\documentstyle[pra,preprint,psfig,aps]{revtex}

\begin{document}
\draft
\title{Dynamical localization and signatures of classical phase space}
\author{Farhan Saif\thinspace\cite{far}}
\address{Department of Electronics, Quaid-i-Azam University, 54320, Islamabad,Pakistan%
\\
and\\
Abteilung f\"ur Quantenphysik, Universit\"at Ulm, Albert-Einstein-Allee 11,
89081 Ulm, Germany.}
\maketitle

\begin{abstract}
We study the dynamical localization of cold atoms in Fermi accelerator both
in position space and in momentum space. We report the role of classical
phase space in the development of dynamical localization phenomenon. We
provide set of experimentally assessable parameters to perform this work in
laboratory.
\end{abstract}

\pacs{PACS numbers: 72.15.Rn, 47.52.+j, 03.75, 03.65.-w}


\section{Introduction}

\label{intro}

Existence of dynamical localization in a system is considered as signature
of quantum chaology~\cite{kn:haak}. Rapid developments in atom optics~\cite
{kn:mlyn,kn:pill,kn:arim} have made this subject a testing ground for the
dynamical localization and hence for quantum chaology. Atomic dynamics in
periodically driven systems, such as, an hydrogen atom in micro-wave field~%
\cite{kn:arnd,kn:bayf,kn:galv} an atom in modulated standing wave field~\cite
{kn:grah,kn:bard,kn:moor}, and the motion of an ion in Paul trap in presence
of standing wave\cite{kn:moham,kn:ried,kn:gardin}, have manifested the
phenomenon of dynamical localization. Latest work on the dynamics of an atom
in Fermi accelerator~\cite{kn:saif1,kn:saift,kn:saif2} has established the
presence of dynamical localization in the system. In addition, this work has
brought into light a new generic phenomenon of dynamical revivals of quantum
chaology~\cite{kn:saift}. In this paper we study Fermi accelerator in atom
optics domain and explain the role of classical chaology in the development
of quantum dynamical localization.

\section{Atomic Fermi Accelerator}

At the end of the first half of twentieth century, Enrico Fermi coined the
idea that the origin and acceleration of cosmic rays is due to intragalactic
giant moving magnetic fields~\cite{kn:fermi}. Latter, Pustilnikov proved
that a particle bouncing on an oscillating surface in gravitational field
may get unbounded acceleration depending upon its initial location in phase
space~\cite{kn:pust}. Based on this idea we have suggested Fermi accelerator
for atoms in atom optics domain and have studied accelerating modes~\cite
{kn:saift}.

We may understand the atomic Fermi accelerator as: Consider a cloud of
cesium atoms initially cold and stored in a magneto-optical trap. On
switching off the trap, the atoms move along the $\tilde{z}$-direction under
the influence of gravitational field and bounce off an atomic mirror~\cite
{kn:wallis,kn:amin}. The latter result from the total internal reflection of
a laser beam incident on a glass prism. The incident laser beam passes
through an acusto-optic modulator which provides a phase modulation to the
evanescent field on the surface of glass prism~\cite{kn:sten}. This model
provides experimental realization of Fermi accelerator in the atom optics
domain~\cite{kn:saif1}.

In order to avoid spontaneous emission we consider a large detuning between
the laser light field and the atomic transition frequency. In presence of
rotating wave approximation and dipole approximation the center-of-mass
motion of the atom in ground state follows from the Hamiltonian 
\begin{equation}
\tilde H \equiv \frac{p^2}{2m} + mg{\tilde z} + \frac{\hbar \Omega_{eff}}{4}
e^{-2k\tilde z+ \epsilon\sin \omega t}
\end{equation}
Here, $\tilde p$ is the momentum of the atom of mass $m$ along the $\tilde{z}
$-axis, $g$ denotes the gravitational acceleration, $\Omega_{eff}$ is the
effective Rabi frequency. The time dependent term expresses the spatial
modulation of amplitude $\epsilon$ and frequency $\omega$.

We introduce the dimensionless position and momentum coordinates $z\equiv 
\tilde{z}\omega ^{2}/g$ and $p\equiv \tilde{p}\omega /(mg)$ and time $%
t\equiv \omega \tilde{t}$. Using these dimensionless coordinates we may
express the Hamiltonian as 
\begin{equation}
H\equiv \frac{p^{2}}{2}+z+V_{0}e^{-\kappa (z-\lambda \sin t)},  \label{dham}
\end{equation}
where, we express the dimensionless intensity $V_{0}\equiv \hbar \omega
^{2}\Omega _{eff}/(4mg^{2})$, steepness $\kappa \equiv 2kg/\omega ^{2}$ and
the modulation amplitude $\lambda \equiv \omega ^{2}\epsilon /(2kg)$ of the
evanescent wave. The commutation relation $[z,p]=[\tilde{z},\tilde{p}]\omega
^{3}/(mg^{2})={\it i}\hbar \omega ^{3}/(mg^{2})$ provides us the
dimensionless Planck's constant $k^{\hspace{-2.1mm}-}\equiv \hbar \omega
^{3}/(mg^{2})$.

The quantum dynamics of atom in Fermi accelerator~\cite{kn:flat1} manifests
dynamical localization in a certain localization window on modulation
amplitude, $0.24<\lambda <\sqrt{k^{\hspace{-2.1mm}-}}/2$~\cite
{kn:saif1,kn:saift}. The lower limit is obtained by Chirikov mapping and
describes the onset of classical diffusion~\cite{kn:chir1,kn:lieb}. The
upper limit of the localization window describes the phase transition of the
quasi-energy spectrum of the Floquet operator from a point spectrum to a
continuum spectrum\cite{kn:benv,kn:oliv,kn:bren,kn:chen}. Above this limit
quantum diffusion sets in and destroys quantum localization. The conditions
of classical and quantum diffusion, together, define the localization window.

\section{Dynamical localization verses the classical phase space}

\label{dlcl}

In order to understand the effect of initial conditions on localization it
is essential to understand how the classical phase space contributes towards
the phenomenon of localization. We find that phase space structure of the
system has direct effect on quantum evolution. In order to study quantum
dynamics within the localization regime, we propagate an initial atomic
wavepacket $\psi(z)$, expressed as 
\begin{equation}
\psi(z)=\frac{1}{\sqrt{\sqrt{2\pi}\Delta z} } \exp\left(-\frac{(z-z_0)^2}{%
2\Delta z^2}\right) \exp\left(-i\frac{p_0 z}{k^{\hspace{-2.1mm}-}} \right),
\end{equation}
at $t=0$, and propagate it in the atomic Fermi accelerator. Here, $z_0$
describes the average position, and $p_0$ denotes the average momentum of
the wave packet. The widths of the wavepacket in position space and in
momentum space are chosen such that they satisfy the minimum uncertainty
condition.

We investigate the effect of classical resonances on dynamical localization
in Fermi accelerator by propagating an atomic wavepacket from an initial
height $z_{0}=20$, with initial momentum $p_{0}=0$. We select $k^{\hspace{%
-2.1mm}-}=1$ which provides localization window on modulation strength, $%
\lambda $, as $0.24<\lambda <0.5$. The initial widths of the wave packet are 
$\Delta z=0.5$ in position space, and $\Delta p=1$ in momentum space,
corresponding to the minimum uncertainty parameters. We propagate the atomic
wavepacket for a modulation amplitude of $\lambda =0.4$ which lies well
within the localization window. We note the probability distribution of the
wavepacket in the atomic Fermi accelerator after evolution $t=1000$, both in
position space and in momentum space. We observe the classical phase space
by means of Poincare' surface of section for the modulation strength $%
\lambda =0.4$, as shown in Fig.~\ref{fg:disph}.

So far as modulation strength is small, that is, $\lambda <\lambda _{l}=0.24$%
, we have isolated resonances in classical phase space. In this domain,
quantum dynamics mimics the classical dynamics and we do not find dynamical
localization. The phenomenon of localization occurs after the overlap of
resonances has occurred in the classical phase space, that is, above $%
\lambda _{l}$ and persists until the quantum diffusion starts in the system~%
\cite{kn:saift}.

Within the localization window we calculate the quantum mechanical position
and momentum distributions of the atomic wavepacket. We compare our result
with the classical phase space observed as Poincare' surface of section. We
place the atomic wavepacket close to the second resonance. Therefore, we
find that maximum probability density is localized there. Hence, a plateau
structure occurs in the probability distribution both in position space and
in momentum space which is at the second resonance of the phase space. Our
numerical results show that the size of the plateau is equal to the size of
the resonance. The tail of the initial Gaussian wavepacket falls
exponentially into the phase space, therefore, it also occupies the other
resonances but with the difference of orders of magnitude. The location of
the first resonance is the closest to the second one, as we find from
Poincare' section of the Fermi accelerator in Fig.~\ref{fg:disph}. As a
result a significant part of initially propagated atomic wavepacket lies in
this region, and seems to be contributing to the plateau structure of second
resonance. We observe that the next plateau corresponding to third primary
resonance is approximately four orders of magnitude smaller, and the next to
it corresponding to the forth resonance of phase space is further eight
order of magnitude smaller. This helps us to infer that the atomic
probability densities in position space and in momentum space localized into
the regions of islands and the atomic wavepacket spreads over the stable
island, making a plateau structure. Outside the stable islands the
probability densities fall linearly into the stochastic sea.

Fishman {\it et. al.}~\cite{kn:fish} have suggested that the eigen functions
decay as $\exp(-(n-{\bar n})/{\ell})$ away from the mean level $\bar n$, in
the momentum space. Therefore, we expect that the overall drop of
probability distribution in momentum space is linear. In momentum space, our
numerical experiment display the overall linear drop of probability
distribution, whereas, in position space the probability distribution
displays an overall drop according to square root law~\cite
{kn:saif1,kn:saift}.

Hence within localization window the atomic wavepacket displays three
interesting features: ({\it i}) {\it plateau structures} in regions
corresponding to stable islands of phase space; ({\it ii}) {\it linear decay}
in regions corresponding to stochastic sea; ({\it iii}) {\it overall decay}
following square root law in position space and following linear behavior in
momentum space. This overall decay may be different for different systems.

We may understand this effect by relating the underlying energy spectrum of
quantum dynamical system with the classical phase space. We~\cite{kn:saifPRE}
have tested that corresponding to a classical resonances there exist a local
discrete spectrum, whereas, in the stochastic region we find quasi
continuum. For the reason we observe that the probability distribution
occupying local discrete spectrum of a resonance undergoes constructive
interference and displays plateau structure, whereas the probability
distribution occupying the quasi continuum spectrum undergoes destructive
interferences and therefore falls linearly in phase space.

\section{Dynamical Localization and changing Panck's constant}

How the change in the effective Planck's constant effects the dynamical
localization? In order to answer the question we propagate the atomic
wavepacket in the Fermi accelerator considering the Planck's constant $k^{%
\hspace{-2.1mm}-}=4$ and compare it with $k^{\hspace{-2.1mm}-}=1$ case. We
keep all the parameters the same as earlier. Following the previous
procedure we note the probability distributions after an evolution time $%
t=1000$ and display it in Fig.~\ref{fg:k14}.

We note that the size of the initial minimum uncertainty wavepacket is
larger due to the larger value of $k^{\hspace{-2.1mm}-}$. Therefore, amount
of the initial probability density falling into the stable islands becomes
larger, as compared with the $k^{\hspace{-2.1mm}-}=1$ case, increasing the
height of the plateaus.. Moreover, for larger $k^{\hspace{-2.1mm}-}$ the
exponential tail of initial Gaussian wavepacket covers more resonances
leading to more plateau structures in the localization arm. The size of the
plateau corresponds to the size of resonances which are independent of the
value of Planck's constant. As a result we conjecture that the size of the
plateau remains the same for two different values of the Planck's constants.

Hence, for the larger value of the Planck's constant, we find that: {\it (i)}
the size of the plateau in probability distributions is the same as it was
for $k^{\hspace{-2.1mm}-}=1$; {\it (ii)} heights of the plateaus are higher; 
{\it (iii)} more plateau structures are appearing.

Comparison between classical and quantum position distributions show that
the plateau structures also appear in the classical cases. However, their
height is larger than the corresponding quantum cases. In order to compare
classical and quantum position distributions we propagate a classical
ensemble. We note the classical distribution (solid thick line) in the Fermi
accelerator after an evolution time $t=1000$ and compare with the
corresponding quantum mechanical distributions for $k^{\hspace{-2.1mm}-}=1$
(solid thin line) and for $k^{\hspace{-2.1mm}-}=4$ (dashed line).

The classical distributions in momentum space and in position space are
entirely different from their quantum counterparts. The classical dynamics
of the ensemble in Fermi accelerator model supports an overall quadratic
distribution in momentum space supporting diffusive dynamics and linear
distribution in position space~\cite{kn:saif1,kn:saift}. Since the classical
position and momentum distributions are the marginal integrations of phase
space, we find that plateaus exist even in classical distributions. We find
that the location and the size of the plateaus are the same in both
classical and quantum cases, however, their heights differ. As compared to
the corresponding classical counterparts, in the quantum mechanical case the
heights of the plateaus is reduced, which may occur as a result of the
dynamical tunneling of the probability to the other plateaus.

\section{Experimental Parameters}

The reflection of atoms onto an evanescent wave mirror has been observed in
many laboratory experiments~\cite{kn:amin,kn:sten,kn:ovch}. In this section
we connect our choice of parameters with the currently accessible technology
and show that the effects we have predicted in this paper can be observed in
a real experiment. We consider cesium atoms of mass $m=2.21\times 10^{-25}$%
kg bouncing on the evanescent wave field with the decay length $%
k^{-1}=0.455~\mu $m and the effective Rabi frequency $\Omega _{eff}=2\pi
\times 5.9$~kHz. These parameters in presence of a modulation frequency of $%
\omega =2\pi \times 1.477$~kHz, lead to $k^{\hspace{-2.1mm}-}=4$, $\kappa
=0.5$ and $V_{0}=4$. By choosing $\Omega _{eff}=2\pi \times 14.9$~kHz, $%
k^{-1}=1.148~\mu $m and $\omega =2\pi \times 0.93$~kHz we get $k^{\hspace{%
-2.1mm}-}=1$, keeping $\kappa =0.5$ and $V_{0}=4$ which are the values used
in our calculations.

\section{Acknowledgment}

\label{ack}

We thank G. Alber, I. Bialynicki-Birula, M. Fortunato, M. El. Ghafar, R.
Grimm, B. Mirbach, M. G. Raizen, V. Savichev, W. P. Schleich, F. Steiner and
A. Zeiler for many fruitful discussions.

%

\begin{figure}[tbp]
\caption{A comparison between classical phase space and quantum mechanical
distributions: At the top we show the Poincar\`{e} surface of section for a
modulation strength $\lambda =0.4>\lambda _{l}$. In (a) we display the
quantum mechanical momentum and in (b) the corresponding position
distribution after a propagation time $t=1000$ for $k^{\hspace{-2.1mm}-}=1$,
using the same value of modulation strength, on logarithmic scales.
Comparing the probability distributions with the classical phase space we
clearly find the probability confinement in the region of a resonance.
However, the probability distribution decays exponentially into the
stochastic region.}
\label{fg:disph}
\end{figure}

\begin{figure}[tbp]
\caption{Change in the position probability distribution with increasing
Planck's constant: We display the probability distribution in position space
for $k^{\hspace{-2.1mm}-}=4$. All the other parameters are the same as in
Fig.~\ref{fg:disph}. We find that with increasing the effective Planck's
constant the height of the plateaus rises indicating an increase in the
probability distribution. However, their location and size remain
approximately the same due to the fixed size of the resonance area. }
\label{fg:k14}
\end{figure}

\begin{figure}[tbp]
\caption{Plateau structures in the classical and in the quantum mechanical
position distributions: We observe that the plateau structures also exist in
the classical distribution. Their location and size are the same as in
quantum distributions, however, they exhibit larger heights. This implies
that, in the classical case, the trapped probability distribution is larger
as compared to the corresponding quantum distribution. A decay of
probability in quantum case may result due to dynamical quantum tunneling
which appears only in quantum cases. In the classical case we propagated
60000 atoms and noted their distribution after $t=1000$, whereas, in our
quantum calculation we followed the same procedure as in Fig.~\ref{fg:disph} 
}
\label{fg:cq}
\end{figure}


\begin{references}
\bibitem[*]{far}  E-mail: saif@physik.uni-ulm.de

\bibitem{kn:haak}  F. Haake, {\it Quantum signatures of Chaos}, (Springer,
Berlin 1992).

\bibitem{kn:mlyn}  J. Mlynek, V. Balykin, and P. Meystre, {\it Special issue
on Optics and Interferometry with Atoms}, Appl. Phys. B {\bf 54}, 319 (1992).

\bibitem{kn:pill}  P. Pillet, {\it Special issue on Optics and
Interferometry with Atoms}, J. Phys. II {\bf 4}, 1877 (1994).

\bibitem{kn:arim}  E. Arimondo and H.A. Bachor, (eds.) {\it Special issue on
Atom Optics}, J. Quant. Semicl. Opt. {\bf 8}, 495 (1996).

\bibitem{kn:arnd}  M. Arndt, A. Buchleitner, R.N. Mantegna, and H. Walther
Phys. Rev. Lett. {\bf 67}, 2435 (1991).

\bibitem{kn:bayf}  J.E. Bayfield, G. Casati, I. Guarneri, and D.W. Sokol,
Phys. Rev. Lett. {\bf 63}, 364 (1989).

\bibitem{kn:galv}  E.J. Galvez, B.E. Sauer, L. Moorman, P.M. Koch, and D.
Richards, Phys. Rev. Lett. {\bf 61}, 2011 (1988).

\bibitem{kn:grah}  R. Graham, M. Schlautmann, and P. Zoller, Phys. Rev. A 
{\bf 45}, R19 (1992).

\bibitem{kn:bard}  P.J. Bardroff, I. Bialynicki-Birula, D.S. Kr\"{a}hmer, G.
Kurizki, E. Mayr, P. Stifter, and W.P. Schleich, Phys. Rev. Lett. {\bf 74},
3959 (1995).

\bibitem{kn:moor}  F.L. Moore, J.C. Robinson, C. Bharucha, P.E. William, and
M.G. Raizen, Phys. Rev. Lett. {\bf 73}, 2974 (1994).

\bibitem{kn:moham}  M. El-Ghafar, P. T\"{o}rm\"{a}, V. Savichev, E. Mayr, A.
Zeiler, and W.P. Schleich, Phys. Rev. Lett. {\bf 78}, 4181 (1997).

\bibitem{kn:ried}  K. Riedel, P. T\"{o}rm\"{a}, V. Savichev, and W.P.
Schleich, Phys. Rev. A {\bf 59}, 797 (1999).

\bibitem{kn:gardin}  S.A. Gardiner, J.I. Cirac, and P. Zoller, Phys. Rev.
Lett. {\bf 79}, 4790 (1997); {\it ibid} Phys. Rev. Lett. {\bf 80}, 2968
(1998). 

\bibitem{kn:saif1}  F. Saif, I. Bialynicki-Birula, M. Fortunato, and W.P.
Schleich, Phys. Rev. A 4779, {\bf 58} 1998.

\bibitem{kn:saift}  F. Saif, Ph.D Thesis on {\it Dynamical Localization and
Quantum Revivals in Driven Systems} (Ulm universit\"{a}t, Ulm, Germany).

\bibitem{kn:saif2}  F. Saif., Phys. Rev. E, accepted for publication.

\bibitem{kn:fermi}  E. Fermi, Phys. Rev. {\bf 75}, 1169 (1949).

\bibitem{kn:pust}  L. D. Pustylnikov, Trans. Moscow Math. Soc. {\bf 2}, 1
(1978).

\bibitem{kn:wallis}  H. Wallis, J. Dalibard, and C. Cohen-Tannoudji, Appl.
Phys. B {\bf 54}, 407 (1992).

\bibitem{kn:amin}  C. G. Aminoff, A. M. Steane, P. Bouyer, P. Desbiolles, J.
Dalibard, C. Cohen-Tannoudji, 
Phys. Rev. Lett. {\bf 71}, 3083 (1993).

\bibitem{kn:sten}  A. Steane, P. Szriftgiser, P. Desbiolles, and J.
Dalibard, Phys. Rev. Lett. {\bf 74}, 4972 (1995).

\bibitem{kn:flat1}  M. Holthaus and M. E. Flatt\'{e}, Phys. Lett. A {\bf 181}%
, 151 (1994).

\bibitem{kn:chir1}  B.V. Chirikov, {\it Phys. Rep.} {\bf 52}, 263 (1979).

\bibitem{kn:lieb}  A.J. Lichtenberg and M.A. Lieberman, {\it Regular and
Stochastic Motion}, (Springer, Berlin, 1983).

\bibitem{kn:benv}  F. Benvenuto, G. Casati, I. Guarneri, and D.L.
Shepelyansky, Z. Phys. B {\bf 84}, 159 (1991).

\bibitem{kn:oliv}  C.R. de Oliveira, I. Guarneri, and G. Casati, Europhys.
Lett. {\bf 27}, 187 (1994).

\bibitem{kn:bren}  N. Brenner and S. Fishman, Phys. Rev. Lett. {\bf 77},
3763 (1996).

\bibitem{kn:chen}  Wen-Yu Chen and G.J. Milburn, Phys. Rev. E {\bf 56}, 351
(1997).

\bibitem{kn:fish}  S. Fishman, D.R. Grempel, and R.E. Prange, Phys. Rev.
Lett. {\bf 49}, 509 (1982).

\bibitem{kn:saifPRE}  F. Saif, G. Alber, V. Savichev, and W. P. Schleich (to
be published).

\bibitem{kn:clem}  See for example, C. Leichtle, I.Sh. Averbukh, and W.P.
Schleich Phys Rev. Lett {\bf 77}, 3999 (1996); Phys Rev. A {\bf 54}, 5299
(1996).

\bibitem{kn:chir2}  B.V. Chirikov and D. Shepelyansky, {\it Radiofizika} 
{\bf 29}, 1041 (1986).

\bibitem{kn:ovch}  Yu.B. Ovchinnikov, I. Manek, and R. Grimm, Phys. Rev.
Lett. {\bf 79}, 2225 (1997).
\end{references}
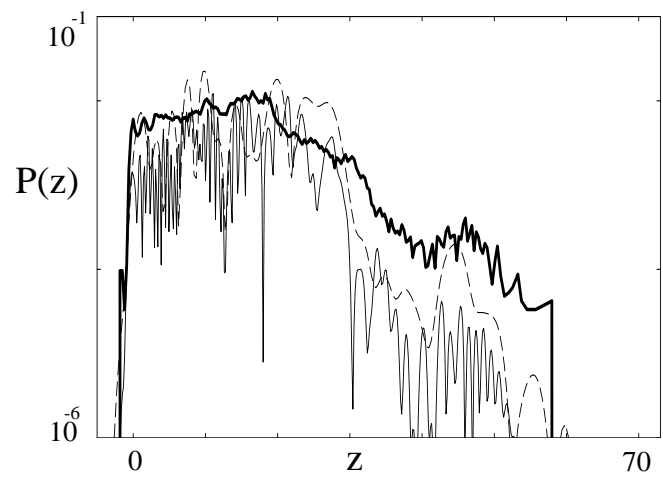
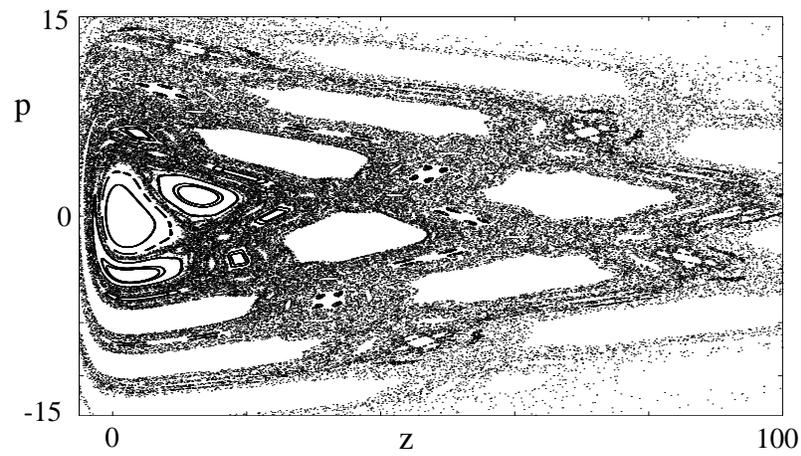
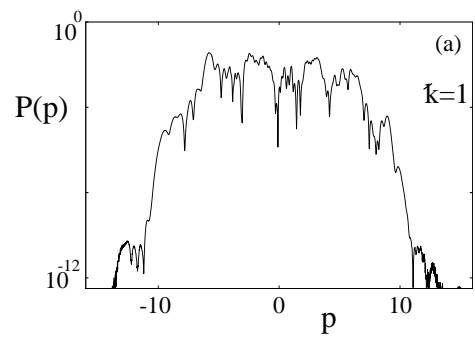
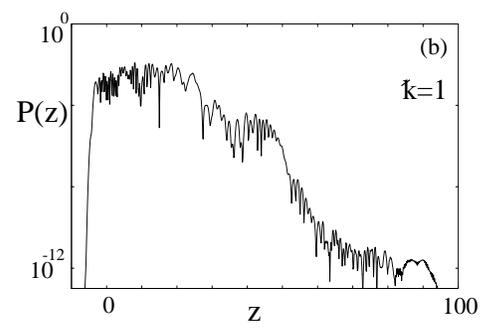
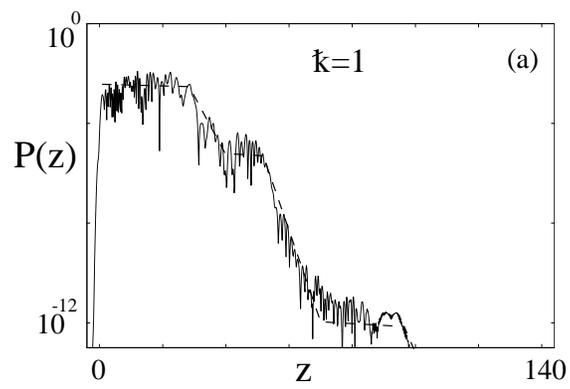
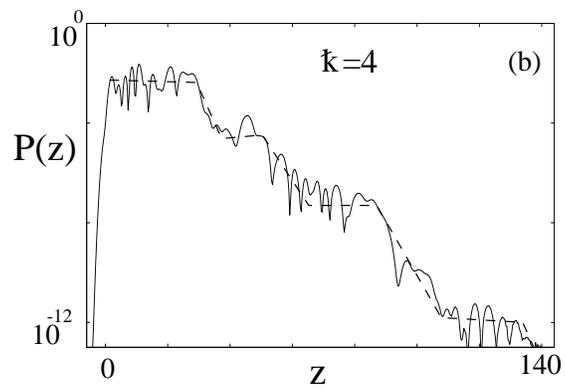
\end{document}